\documentclass[aps,pra,letterpaper,10pt,twocolumn,showpacs,superscriptaddress]{revtex4-1}
\usepackage{hyperref}
\usepackage{amssymb}
\usepackage{amsmath}
\usepackage{graphicx}
\usepackage{subfigure}
\usepackage{amsfonts}
\usepackage{color}
\newcommand{\rmc}{\mathrm{c}}
\newcommand{\rmd}{\mathrm{d}}
\newcommand{\rmm}{\mathrm{m}}
\newcommand{\rmD}{\mathrm{D}}
\newcommand{\rmL}{\mathrm{L}}
\newcommand{\rmin}{\mathrm{in}}

\newcommand{\bfu}{\boldsymbol{u}}
\newcommand{\bfn}{\boldsymbol{n}}

\newcommand{\etamin}{\eta_{\mathrm{min}}}
\newcommand{\trp}{\mathsf{T}}

\newcommand{\mean}[1]{\overline{#1}}
\begin{document}
\title{Entangling the motion of two optically trapped objects via time-modulated driving fields}

\author{Mehdi Abdi}
\affiliation{Technische Universit{\"a}t M{\"u}nchen, Physik Department, James Franck Stra{\ss}e, 85748 Garching, Germany}
\author{Michael J. Hartmann}
\affiliation{Institute of Photonics and Quantum Sciences, Heriot-Watt University, Edinburgh, EH14 4AS, United Kingdom}

\date{\today}
\begin{abstract}
We study entanglement of the motional degrees of freedom of two tethered and optically trapped microdisks inside a single cavity.
By properly choosing the position of the trapped objects in the optical cavity and driving proper modes of the cavity it is possible to equip the system with linear and quadratic optomechanical couplings.
We show that a parametric coupling between the fundamental vibrational modes of two tethered mircodiscs can be generated via a time modulated input laser.
For a proper choice of the modulation frequency, this mechanism can drive the motion of the microdisks into an inseparable state in the long time limit via a two-mode squeezing process.
We numerically confirm the performance of our scheme for current technology and briefly discuss an experimental setup which can be employed for detecting this entanglement by employing the quadratic coupling. We also comment on the perspectives for generating such entanglement between the oscillations of optically levitated nanospheres.
\end{abstract}

\maketitle

%
%
\section{Introduction}\label{sec:introduction}
Entanglement, despite its wide variety of proposed potential applications, has been a challenging feature in theory of quantum mechanics from its first days of advent \cite{Einstein1935}.
Besides its fundamental significance, its prominent role in future information technology makes it a concept with large number of attendees.
Another intensively debated aspect of quantum physics is its border with classical physics,
and how a quantum state of a specific system passes to a classical state due to decoherence introduced to the system as a result of its interaction with the remainder of the world, the environment \cite{Zurek2003}.
In attempts to understand these connections, speculations on whether quantum mechanics would eventually cease to be valid at macroscopic scales have been put forward \cite{Adler2009}.
Investigating the generation of distinct quantum mechanical, i.e. non-classical, states for ever larger or eventually \emph{macroscopic} objects is thus important for improving our understanding of the quantum to classical transition.
This could be done by creating quantum states such as superposition states \cite{Romero-Isart2011,Pepper2012}, squeezed states \cite{Nunnenkamp2010,Lue2014}, or Fock states \cite{Rips2012,Rips2013,Rips2014} in macroscopic objects.
Hence, bringing the two concepts, entanglement and quantum behavior of macroscopic objects, together and making micro--macro entangled states \cite{Ghobadi2014} or entangling two macroscopic systems \cite{Pirandola2006,Hartmann2008,Abdi2012} in order to study their evolution and transition to a separable classical state can shed more light on debated aspects of quantum mechanics.
Optomechanical systems are one of the candidates for realizing such investigations \cite{Aspelmeyer2014}.

Entangling two mechanical resonators in optomechanical systems calls for high mechanical quality factors.
These can be achieved by reducing interactions between the mechanical resonators and their thermal environment, a task which brought many efforts into laboratories.
Here, strong coupling of light with matter that exceeds dissipative processes can lead to optomechanical entanglement \cite{Vitali2007,Palomaki2013}.
One approach to achieve high quality mechanical oscillations that has been proposed \cite{Chang2010,Romero-Isart2010,Lechner2013} and implemented \cite{Kiesel2013,Millen2014a} during recent years is to optically trap dielectric objects as a mechanical component for an optomechanical system.
However such levitated nano-objects suffer decoherence due to the recoils of photons from the trapping fields.
In particular spheres can scatter recoil photons into the entire solid angle. A way to suppress this decoherence mechanism can be to arrange for a situation where the majority of recoil photons need to be scattered into a discrete set of modes, e.g. the resonant modes of a high finesse optical cavity.   
Thus, alternative setups have been proposed and implemented which have both high quality factors and high cooperativity \cite{Chang2012,Ni2012}.
Dielectric tethered microdisks, for example, have proven to be capable of functioning as sensors for detecting high-frequency gravitational waves \cite{Arvanitaki2013}.
Furthermore, they in essence have the same abilities as optically levitated dielectric objects with respect to cavity cooling \cite{Schulze2010,Pender2012}, generating optomechanical entanglement \cite{Nie2012}, and preparing them in quantum superposition states \cite{Romero-Isart2011}.

Motivated by both, technical applications and the fundamental importance of generating purely mechanical entangled states, we exploit the advantages of such systems to study entanglement properties of two optically trapped dielectric objects inside a single cavity.
Taking into account the relevant harmful sources of decoherence, like photon recoil heating and impact of air molecules,
we will see that mechanical resonators enhanced by optical trapping, like tethered microdisks \cite{Ni2012}, are appropriate candidates for creating such entangled states.
We find that, although the steady state of a continuously driven system composed of two tethered membranes and several cavity modes does, for realistic parameters, not show any mechanical--mechanical entanglement, it can become inseparable by introducing a parametric interaction into the mechanical system \cite{Galve2010} via a time-modulated input laser.
An analytical treatment of the system with an adiabatic elimination technique suggests to use modulated input lasers to excite a two-phonon squeezing process which brings in a fully mechanical entangled Gaussian state.
We also find that it is rather challenging to create purely mechanical entanglement between two fully levitated nanoparticles as their size crucially determines their coupling to optical cavity modes and the amount of scattered cavity photons.
In fact, this trade off propels one to smaller particles leading to smaller coupling rates.

The remainder of the paper is organized as follows:
In Sec.~\ref{sec:model} the general model and Hamiltonian of the system is introduced.
We will then study the dynamics of the system in Sec.~\ref{sec:dynamics}.
Steady states of the system are considered in Sec.~\ref{sec:steady-state}.
Sec.~\ref{sec:parametric} is then devoted to discussions of the parametric coupling of the mechanical resonators and entangling their motional degrees of freedom.
We will also briefly discuss an entanglement measuring method.
Some considerations about entanglement of two levitated nanospheres are presented in Sec.~\ref{sec:nanospheres}.
Finally, a summary of the paper and concluding remarks are brought in Sec.~\ref{sec:conclusion}.

%
%
\section{The model}\label{sec:model}
We consider a Fabry--Perot cavity where the dielectric tethered membranes are mounted in.
Each microdisk is suspended from its support by a narrow band which loosely confines its motion.
These dielectric objects are then subjected to an optical field and experience an electric force via a dipole interaction with the light resulting in an enhanced (both in frequency and quality factor) mechanical resonator.
Here, we focus on a 1D model where three longitudinal cavity modes are fed by three lasers at frequencies $\omega_{\rmL,i}$ with $i=0,1,2$.
In our model one of the cavity modes is strongly driven and will act as a trap for the microdisks while the two remaining modes provide the optomechanical coupling.
The objects are trapped at the anti-nodes of the $i=0$ cavity mode, the one which hereafter will be called the trapping mode.
However, the two other optical modes will also contribute to the trapping and will \textit{slightly} modify the trap frequency and shift the equilibrium position of the tethered membranes.
The Hamiltonian of the system ,in a frame rotating with the input laser frequencies, is given by \cite{Chang2012}
\begin{eqnarray}
\hat{H} &=& \sum_{i=0}^{2}\hbar\delta_{i}\hat{a}_{i}^{\dagger}\hat{a}_{i} 
+\sum_{j=1}^{2} \frac{\hat{p}_{j}^{2}}{2m_{j}} + i \sum_{i=0}^{2}\hbar E_{i}(t)(\hat{a}_{i}^{\dagger}-\hat{a}_{i}) \nonumber\\
&&-\sum_{i,j}\hbar g_{ij}\hat{a}_{i}^{\dagger}\hat{a}_{i}\cos^{2}(k_{i}\hat{x}_{j}-\phi_{ij}) ,
\label{hamiltonian}
\end{eqnarray} 
where $\delta_{i}=\omega_{i}-\omega_{\rmL,i}$ is the detuning of the laser from the cavity resonance, whose field is described by annihilation (creation) operators $\hat{a}_{i}$ ($\hat{a}_{i}^{\dagger}$) satisfying the commutation relation $[\hat{a}_{i},\hat{a}_{i}^{\dagger}]=1$.
Moreover, $k_{i}=2\pi/\lambda_{i}$ is wave number of each cavity mode and $E_{i}$ is a measure of the $i$th cavity field amplitude.
In (\ref{hamiltonian}) the mechanical objects, are identified by their position $\hat{x}_{j}$ and momentum $\hat{p}_{j}$ with commutator $[\hat{x}_{j},\hat{p}_{j}]= i \hbar$.
The interaction of the dielectric objects with the cavity fields is provided by a dipole force which leads to both, their optical trapping and their optomechanical coupling to the cavity field.
The polarizability of the microdisks determines the coupling factor; if $V_{\rmc,i}$ is the volume of the $i$th cavity mode, $V_{\rmd,j}$ the volume of the $j$th dielectric object, and $\epsilon$ its relative permittivity, the optomechanical coupling reads $g_{ij}=\frac{V_{\rmd,j}}{2V_{\rmc,i}}(\epsilon -1)\omega_{i}$.

We note that the model presented in this section is general and is applicable to most systems with optically trapped dielectric objects; nanospheres \cite{Chang2010}, microdisks \cite{Chang2012}, and nanodumbbells \cite{Lechner2013}.
However, we will concentrate on the case of tethered microdisks and will shortly discuss a version with nanospheres in section \ref{sec:nanospheres}.

In order to get stably trapped objects, we assume that the trapping cavity mode is intensively driven such that the motion of the microdisks is well localized at its antinodes, i.e., $\phi_{0j}=0$ in (\ref{hamiltonian}).
To keep this valid even in the presence of the control cavity modes ($i=1,2$), one needs to assure that the intracavity intensity of the trap field is much higher than for the other modes $E_{0}\gg E_{1}, E_{2}$.
Hence, the trapped objects are in the Lamb--Dicke regime with $k_{i}\langle \hat{x}_{j} \rangle \ll 1$.
These conditions allow us to expand the cosine term in the Hamiltonian and keep up to the terms proportional to $(k_{i}\hat{x}_{j})^{2}$.
We will also see that it is sensible to neglect the quantum fluctuations of the $i=0$ mode and assume its only task is to provide the trap.
Hence, the Hamiltonian in terms of dimensionless mechanical quadratures $\hat{x}_{j}$ and $\hat{p}_{j}$ (with $[\hat{x}_{j},\hat{p}_{j}]= i $) reads
\begin{eqnarray}
\frac{\hat{H}}{\hbar} &=& \sum_{i=1}^{2}\delta_{i}\hat{a}_{i}^{\dagger}\hat{a}_{i} 
+\sum_{j=1}^{2} \frac{\Omega_{j}}{2}(\hat{p}_{j}^{2} +\hat{x}_{j}^{2}) + i \sum_{i=1}^{2} E_{i}(t)(\hat{a}_{i}^{\dagger}-\hat{a}_{i}) \nonumber\\
&& +\sum_{i,j=1}^{2}\hat{a}_{i}^{\dagger}\hat{a}_{i}(\mathcal{G}_{ij}^{\mathrm{l}}\hat{x}_{j} +\mathcal{G}_{ij}^{\mathrm{q}}\hat{x}_{j}^{2}) ,
\label{lambham}
\end{eqnarray}
where the tethered membranes undergo harmonic oscillations with the frequency $\Omega_{j}^{2}=\frac{2\hbar k_{0}^{2}}{m_{j}} g_{0j}|\langle \hat{a}_{0}\rangle|^{2}$ provided by the optical trap.
The linear and quadratic optomechanical coupling constants $\mathcal{G}_{ij}^{\mathrm{l}}$ and $\mathcal{G}_{ij}^{\mathrm{q}}$ are respectively defined as
\begin{eqnarray*}
\mathcal{G}_{ij}^{\mathrm{l}} &=& -\sqrt{2}k_{i}x_{\mathrm{zp},j}g_{ij}\sin(2\phi_{ij}), \\
\mathcal{G}_{ij}^{\mathrm{q}} &=& 2k_{i}^{2}x_{\mathrm{zp},j}^{2}g_{ij}\cos(2\phi_{ij}),
\end{eqnarray*}
where, $x_{\mathrm{zp},j}=\sqrt{\hbar/2m_{j}\Omega_{j}}$ is the zero-point motion of the $j$th dielectric object.
It is important to bear in mind that $\mathcal{G}_{ij}^{\mathrm{q}} \ll \mathcal{G}_{ij}^{\mathrm{l}}$, except for a particle placed at the antinode of a cavity mode.
Note also that the laser detuning in this new form of the Hamiltonian is $\tilde{\delta}_{i}=\omega_{i}-\omega_{\rmL,i}-\sum_{j=1}^{2}g_{ij}\cos^{2}\phi_{ij}$, where we have dropped the tilde in (\ref{lambham}).
We now turn to study dynamics of the above system.

%
%
\section{Dynamics}\label{sec:dynamics}
The full dynamics of the system can conveniently be studied by quantum Langevin equations, which include damping processes acting on the system, the associated noises and other sources of decoherence.
The quantum Langevin equations corresponding to the Hamiltonian (\ref{lambham} are
\begin{subequations}
\begin{eqnarray}
\dot{\hat{x}}_{j} &=& \Omega_{j}\hat{p}_{j}, \\
\dot{\hat{p}}_{j} &=& -\Omega_{j}\hat{x}_{j} -\gamma_{j}\hat{p}_{j} -\sum_{i=1}^{2}\hat{a}_{i}^{\dagger}\hat{a}_{i} (\mathcal{G}_{ij}^{\mathrm{l}} +2\mathcal{G}_{ij}^{\mathrm{q}}\hat{x}_{j}) +\hat{\xi}_{j}, \\
\dot{\hat{a}}_{i} &=& -(\kappa_{i}+ i \delta_{i})\hat{a}_{i} +E_{i}(t) +\sqrt{2\kappa_{i}}\hat{a}_{i}^{\rmin} \nonumber\\
&&-i \sum_{j=1}^{2}\hat{a}_{i}(\mathcal{G}_{ij}^{\mathrm{l}}\hat{x}_{j} +\mathcal{G}_{ij}^{\mathrm{q}}\hat{x}_{j}^{2}),
\end{eqnarray}\label{qles}
\end{subequations}
where $\kappa_{i}$ is cavity decay rate for the $i$th mode which is associated with the input vaccum noise operator $\hat{a}_{i}^{ i n}$.
The decay and decoherence in the cavity modes is dominantly due to leakage through the input mirror and scattering of photons from the edge of the microdisks.
In addition to the tether attached to the microdisks, their mechanical motion is affected by the random impacts of the chamber air molecules which lead to damping of their oscillations at a rate $\gamma_{j}$.
This rate is in direct proportion to the chamber pressure $\mathcal{P}$ and inversely proportional to the mean thermal velocity of the air molecules $\bar{v}$ \cite{Chang2012} ($\bar{v} =\sqrt{3 k_{\mathrm{B}} T/m_{\mathrm{air}}}$ where $T$ is the temperature of the chamber, $k_{\mathrm{B}}$ is Boltzmann's constant, and $m_{\mathrm{air}}$ is the mass of the air molecules.).
However, the coherence of the mechanical motion is mostly affected by fluctuations in the optical trap \cite{Grotz2006} stemming from scattering of the cavity photons by the dielectric object.
In fact, this is the major phenomenon affecting quantum nature of the mechanical oscillators and substantially reduces their cooperativity.
For levitated nanoshperes this decoherence is very destructive as the photons are scattered to any direction, accessing an infinite number of free space modes out of the \emph{effectively} one-dimensional cavity.
However, in the case of a tethered microdisk the effect of scattered photons is much smaller as photons are predominantly scattered in the direction of the cavity access, where they can only be scattered into a small discrete set of cavity modes.
Yet, the scattering of cavity photons out of the cavity also brings in extra cavity decay source modifying the cavity finesse. We denote this modified optical finesse by $\mathcal{F}_{\mathrm{eff}}$.

Since at room temperature the number of thermal optical photons is very small, the only non-zero correlation function for the cavity modes is $\langle \hat{a}_{i}^{ i n}(t)\hat{a}_{i}^{ i n,\dagger}(t')\rangle =\delta(t-t')$.
The Markovian approximation for the fluctuations in mechanical oscillation of the dielectric objects is valid for their relatively low frequencies and one adopts the following correlation function
\begin{equation}
\langle \hat{\xi}_{j}(t)\hat{\xi}_{j}(t') \rangle_{\mathrm{sym}} =(2\bar{n}_{\mathrm{th},j}+1)(\gamma_{j}+\Gamma_{j})\delta(t-t'),
\label{corxi}
\end{equation}
where $\bar{n}_{\mathrm{th},j}=[\exp(\frac{\hbar\Omega_{j}}{k_{\mathrm{B}}T})-1]^{-1}$ is mean phonon number of the $j$th mechanical resonator connected to a thermal bath at equilibrium temperature $T$.
$\Gamma_{j}$ describes the diffusion of the mechanical momentum stemming from photon recoil decoherence.
Neglecting scattering of the control mode photons we have $\Gamma_{j}=\frac{\lambda_{0}}{4L}\frac{V_{\rmc,0}}{V_{\rmd,j}}\frac{\Omega_{j}}{\mathcal{F}_{\mathrm{eff}}(\epsilon -1)}$.

\subsection*{Linearization}
The quantum Langevin equations of (\ref{qles}) are nonlinear, but they can be linearized in the parameter region we are interested in.
Actually, in order to achieve strong optomechanical coupling which is required for attaining stationary entangled states, one needs to intensively drive the cavity modes.
This allows us to suppose that the field inside the cavity is composed of a large coherent part and some quantum fluctuations around this classical state.
Thus, it is valid to transform each operator of the system as $\hat{o} \mapsto \mean{o}+\hat{o}$ and then focus on the fluctuations around coherent part of the variable.
This transformation holds for mechanical oscillators as well, expressing a new equilibrium position for them.
As the quantum fluctuations are very small compared to the classical parts it is then reasonable to neglect any quadratic and higher order terms and only keep the first order terms in $\hat{o}$.
Consequently, there will be no quadratic optomechanical interaction which is a valid approximation as long as $k_{i}\mean{x}_{j} \ll 1$.
We note that there is of course a \textit{pure} quadratic optomechanical interaction which couples dynamics of the trapping cavity field and the dielectric objects.
However, because of the deep optical trapping its effect on the dynamics of the microdisks is much smaller than the linear coupling of the other modes and we have omitted such interactions in treating the system.
Strictly speaking, this approximation is valid when the amplitudes of the cavity modes obey $E_{0}k_{i}x_{\mathrm{zp},j} \ll E_{i}$.
This criterion also allows for leaving out the quantum dynamics of the trapping mode.

The coherent parts of the system variables obey the following dynamics
\begin{subequations}
\begin{eqnarray}
\dot{\mean{x}}_{j} &=& \Omega_{j}\mean{p}_{j}, \\
\dot{\mean{p}}_{j} &=& -\Omega_{j}\mean{x}_{j} -\gamma_{j}\mean{p}_{j} -\sum_{i=1}^{2} G_{ij}\mean{a}_{i}^{*}, \\
\dot{\mean{a}}_{i} &=& -(\kappa_{i}+ i \Delta_{i})\mean{a}_{i} +E_{i}(t),
\end{eqnarray}\label{firstm}
\end{subequations}
where an effective cavity detuning is defined as $\Delta_{i} = \delta_{i} +\sum_{j=1}^{2}(\mathcal{G}_{ij}^{\mathrm{l}}\mean{x}_{j} +\mathcal{G}_{ij}^{\mathrm{q}}\mean{x}_{j}^{2})$ and an effective linear optomechanical coupling strength as $G_{ij} = \mean{a}_{i}(\mathcal{G}_{ij}^{\mathrm{l}}+2\mathcal{G}_{ij}^{\mathrm{q}}\mean{x}_{j})$.
The quantum fluctuations are expressed by a series of linear Langevin equations: 
\begin{subequations}
\begin{eqnarray}
\dot{\hat{x}}_{j} &=& \Omega_{j}\hat{p}_{j}, \\
\dot{\hat{p}}_{j} &=& -\tilde{\Omega}_{j}\hat{x}_{j} -\gamma_{j}\hat{p}_{j} -\sum_{i=1}^{2}(G_{ij}\hat{a}_{i}^{\dagger} +G_{ij}^{*}\hat{a}_{i}) +\hat{\xi}_{j}, \\
\dot{\hat{a}}_{i} &=& -(\kappa_{i}+ i \Delta_{i})\hat{a}_{i} - i \sum_{j=1}^{2}G_{ij}\hat{x}_{j} +\sqrt{2\kappa_{i}}\hat{a}_{i}^{\rmin},
\end{eqnarray}\label{secondm}
\end{subequations}
where $\tilde{\Omega}_{j} =\Omega_{j}+2\sum_{i=1}^{2}|\mean{a}_{i}|^{2}\mathcal{G}_{ij}^{\mathrm{q}}$ indicates a \emph{slight} modification of the trap stiffness by the control cavity modes.

After linearizing the quantum Langevin equations the effective Hamiltonian which would produce these equations is quadratic in the system operators, and therefore, if we start the system in a Gaussian state it will retain its Gaussian nature.
Moreover, all noise operators have zero-mean Gaussian correlations.
Hence, the system is fully characterized by its first moments calculated from (\ref{firstm}) and the second moments that could be obtained from (\ref{secondm}).
We now define Hermitian optical field quadratures $\hat{X}_{i}$ and $\hat{Y}_{i}$ via $\hat{a}_{i} =(\hat{X}_{i}+i\hat{Y}_{i})/\sqrt{2}$.
With the latter, the dynamics can be expressed in the compact form
\begin{equation}
\dot{\hat{\bfu}}=\mathbf{A}\hat{\bfu} +\hat{\bfn},
\label{compact}
\end{equation}
where the operator and noise vectors are defined as
\begin{eqnarray*}
\hat{\bfu} &=&[\hat{x}_{1},\hat{p}_{1},\hat{x}_{2},\hat{p}_{2},\hat{X}_{1},\hat{Y}_{1},\hat{X}_{2},\hat{Y}_{2}]^{\trp}, \\
\hat{\bfn} &=&[0,\hat{\xi}_{1},0,\hat{\xi}_{2},\sqrt{2\kappa_{1}}\hat{X}_{1}^{\rmin},\sqrt{2\kappa_{1}}\hat{Y}_{1}^{\rmin},\sqrt{2\kappa_{1}}\hat{X}_{2}^{\rmin},\sqrt{2\kappa_{1}}\hat{Y}_{2}^{\rmin}]^{\trp}.
\end{eqnarray*}
In general, the drift matrix $\mathbf{A}$ (see Appendix A for its explicit form) is a function of time via $\tilde{\Omega}_{j}$, the detuning parameters $\Delta_{i}$, and the coupling factors $G_{ij}$ because of time dependence of the field amplitudes $E_{i}$.

%
%
\section{Steady state of the system}\label{sec:steady-state}
First let us study the steady state of the system and investigate possible stationary entanglement between motional degrees of freedom of the two distinct, optically trapped dielectric objects.
The state of the whole system is a zero-mean Gaussian state, because we have linearized the dynamics around the first moments.
Therefore, to characterize the state of the system one needs only to compute the covariance matrix of the system.
When the driving lasers are CW, $E_{i}$s are constant, the system will arrive at its steady state provided that it is stable.
The stability of the system can be checked via a Routh--Hurwitz criterion \cite{Gopal2002}.
Then the covariance matrix $\mathbf{V}$ with elements $V_{ij}=\langle\hat{u}_{i}\hat{u}_{j}+\hat{u}_{j}\hat{u}_{i}\rangle/2$ of the steady state can be computed from the following Lyapunov equation \cite{Genes2008}
\begin{equation}
\mathbf{A}\mathbf{V}+\mathbf{V}\mathbf{A}^{\trp}+\mathbf{D} =0,
\label{lyapunov}
\end{equation}
which will, of course, give a time-independent covariance matrix.
In (\ref{lyapunov}), $\mathbf{D}$ is the diffusion matrix (the matrix of noise correlations) given by a diagonal matrix $\mathbf{D} =\mathrm{diag}[0, \frac{2k_{\mathrm{B}}T}{\hbar\Omega_{1}}(\gamma_{1}+\Gamma_{1}), 0, \frac{2k_{\mathrm{B}}T}{\hbar\Omega_{2}}(\gamma_{2}+\Gamma_{2}), \kappa_{1}, \kappa_{1}, \kappa_{2}, \kappa_{2}]$.

In order to achieve quantum states in the system of two mechanical oscillators we need to cool down both, the centre of mass and the breathing mode of the system.
According to the interaction term in (\ref{lambham}) this can be achieved by adjusting the phase factors $\phi_{ij}$.
In fact, the optimal phase values for getting both collective modes cooled down and realizing maximum optomechanical coupling are $\phi_{11}=\pm\phi_{12}=\phi_{21}=\mp\phi_{22}=\frac{\pi}{4}$.
Here we consider these phase values for wavelengths around $\lambda_{i} \approx 1064 \mu\rmm$, which in principle can be obtained by properly positioning the objects inside the cavity.
In practice, it is possible to achieve these phase values by adjusting position of the dielectric objects and driving proper longitudinal cavity modes.
Actually, for the $i$th cavity mode the equation relating phases of two trapped objects is $\phi_{i2}-\phi_{i1} =nk_{i}\lambda_{0}$, where $n \in \mathbb{Z}$.
Hence, the two control parameters are the wavelength ($\lambda_{i}$) and the relative position ($n$th antinode of the trapping mode).
Note also that, to avoid zero total optomechanical coupling for the object with $\phi_{1j}=-\phi_{2j}=\pm\frac{\pi}{4}$ one needs to choose different input powers and/or different cavity detuning, or slightly modify these phase values.
Here, we numerically look for phase values, relative intracavity field amplitudes, and detunings which optimize the measure of the entanglement.
The measure of bipartite entanglement we will use is the logarithmic negativity \cite{Vidal2002,Plenio2005}
\begin{equation}
E_{N} = \mathrm{max}\{0,-\log(2\etamin)\},
\end{equation}
where $\etamin$ is the minimum symplectic eigenvalue of the partially transposed covariance matrix.
Note that a bipartite state is inseparable when $\etamin<\frac{1}{2}$.

\begin{figure}[t]
\centering
\includegraphics[width=\linewidth]{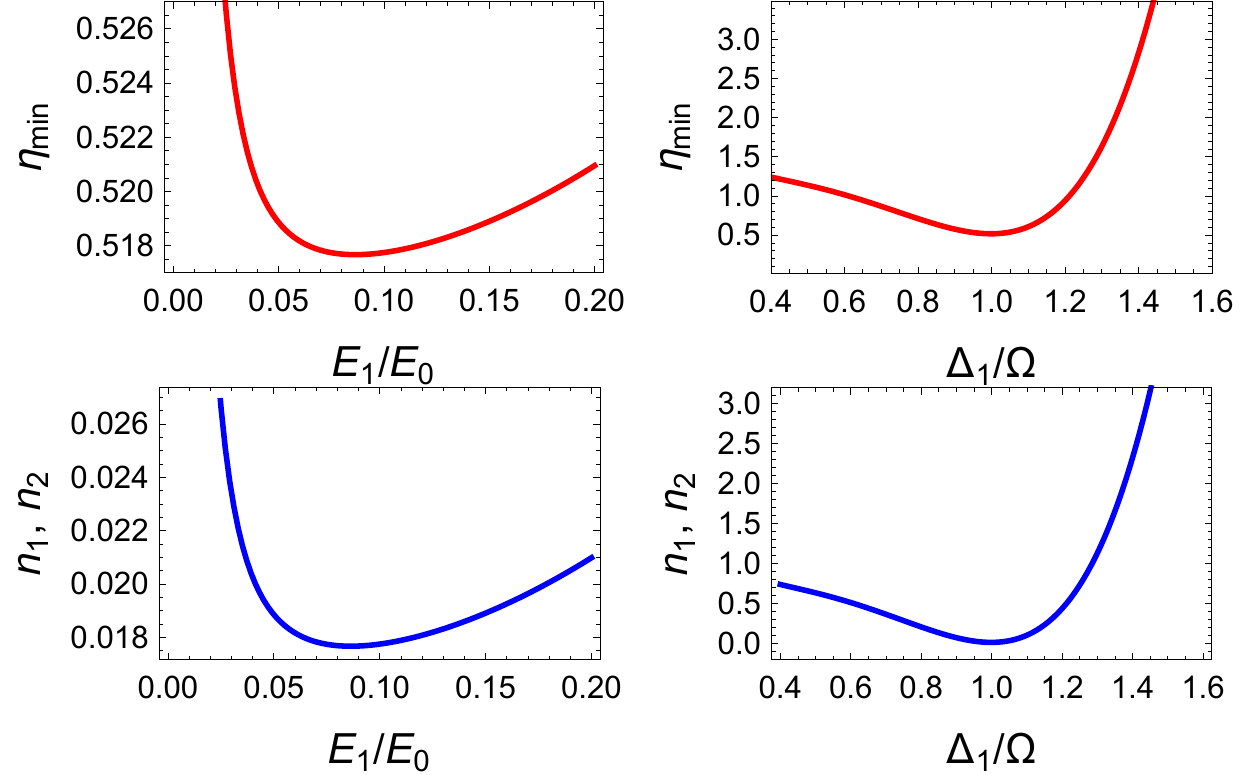}
\caption{(color online).
Variations of $\eta_{\mathrm{min}}$ (upper panels) and mean phonon occupation numbers $\bar{n}_{1}$ and $\bar{n}_{2}$ (lower panels) for two microdisks of $20~\mu\mathrm{m}$ diameter and $150~\mathrm{nm}$ thickness with respect to intracavity amplitudes when $\Delta_{1}=\Delta_{2}=+\Omega$ (left panels) and input laser detuning when $E_{1}=E_{2}=0.1E_{0}$ (right panels).
In the plots we have set $E_{2}=E_{1}$ and $\Delta_{2}=\Delta_{1}$.
To get the oscillation frequency listed in Table~\ref{parameters} one feeds the trapping mode of the cavity by a $15~\mathrm{mW}$ laser.}
\label{micross}
\end{figure}

We first examine the steady state of the system for experimentally feasible parameters.
In Fig.~\ref{micross} variations of $\eta_{\mathrm{min}}$ versus the intracavity amplitudes of the optical control modes and their detuning is plotted for a cavity containing two identical tethered membranes.
Since we are interested in the mechanical--mechanical entanglement, $\etamin$ corresponds to the reduced $4 \times 4$ covariance matrix which only contains mechanical covariances.
It is obvious from the plots that the steady state of the system gets very close to the inseparability verge $\etamin=0.5$.
As we will see in the next section, it is possible to cross the border by establishing a parametric mechanical--mechanical coupling.

We note that in contrast to the membrane in the middle setup \cite{Hartmann2008}, the opomechanical coupling can for optically trapped dielectric objects not be enhanced on a similar scale by increasing the intracavity light intensities. Inseparable mechanical states could in both setups be reached by increasing the effective optomechanical coupling $G_{ij}$, which in practice is done by enhancing the light intensity inside the cavity. However, to avoid wipeout of the stable optical trap one also needs to increase intensity of the trapping mode in expense of increasing decoherence rate stemming from the photon recoil process. This behavior precludes the generation of entanglement between optically trapped objects with constant input fields merely by increasing their intensity.

\begin{table}[tbs]
\caption{\label{parameters}Parameters of the optomechanical system.}
\begin{ruledtabular}
\begin{tabular}{lcc}
\textrm{Quantity} & \textrm{Microdisk} & \textrm{Nanosphere} \\
\colrule
$\Omega/2\pi$ & $11~\mathrm{MHz}$ & $5~\mathrm{MHz}$ \\
$Q_{\rmm}$ & $1\times10^{6}$ & $3\times10^{8}$ \\
$\mathcal{F}_{\mathrm{eff}}$ & $7\times10^{5}$ & $4\times10^{5}$ \\
$m$ & $10^{+2}~\mathrm{pg}$ & $10^{-2}~\mathrm{pg}$
\end{tabular}
\end{ruledtabular}
\end{table}

Fig.~\ref{micross} also shows the steady state mean phonon number of the microdisks $\bar{n}_{j}=(\langle\hat{x}_{j}^{2}\rangle+\langle\hat{p}_{j}^{2}\rangle-1)/2$---which is the same for both of the objects as we have considered them to be identical.
Note that the mode amplitudes at which the $\eta_{\mathrm{min}}$ and $\bar{n}_{j}$ are minimal are not in conflict with our condition for stable trapping ($E_{i} \ll E_{0}$).
One also notices that the optimal values for laser detuning are $\Delta_{i}=\Omega_{1} \approx \Omega_{2}$.
In our numerics we have considered silica dielectric objects with $\epsilon = 2.1$ and $\rho = 2201~\mathrm{Kgm}^{-3}$.
The length of the Fabry--P\'erot cavity is $L=1~\mathrm{mm}$ and the wavelength of all the input lasers is set around $1064~\mathrm{nm}$.
We assume that the pressure of the chamber containing the setup is $\mathcal{P}=10^{-6}~\mathrm{mbar}$,
a value that has already been reached in recent experiments \cite{Ni2012}.
The effective temperature of the system is taken to be $T=100~\mathrm{mK}$ which can be attained by a feedback mechanism for precooling the system from room temperature (it can be a process similar to the technique used in \cite{Gieseler2012} for cooling nanospheres).
We remind that even though the model considered here is fully based on cavity trapping and optomechanics, it is in principle compatible with feedback trapping or optical tweezers setups.
All other relevant parameters are listed in Table~\ref{parameters}.
To consider the effect of the tether attached to the microdisks we take a moderate quality factor.
The quality factor one would get for a levitated microdisk is solely determined by the air molecule impacts, which for pressures as low as $10^{-6}$ mbar is $\sim4\times10^{9}$.
In our considerations $Q_{\rmm}$ is taken to be three orders of magnitude smaller (see Table~\ref{parameters}).
This value is considered to be within reach of moderate experimental improvements \cite{Ni2012}.

%
%
\section{Parametric coupling via time-modulated input fields}\label{sec:parametric}
The failure in creating steady state entanglement between the mechanical motion of optically trapped objects for realistic parameters leads us to an alternative way for attaining it.
Entanglement between two objects can originate from a mutual interaction.
Intuitively, it is clear that any interaction between the mechanical motion of the levitated objects is mediated by the optical cavity fields.
However, to get a more detailed picture about this fully mechanical interaction, we eliminate the role of cavity field dynamics and look only at the mechanical resonators.
That is, we assume that the system operates in the \textit{weak coupling} regime $G_{ij}<\kappa_{i}$, and adiabatically eliminate the cavity modes to arrive at: $\dot{\hat{x}}_{j} =\Omega_{j}\hat{p}_{j}$ and
\begin{eqnarray}
\dot{\hat{p}}_{j} &=-\tilde{\Omega}_{j}\hat{x}_{j} -\gamma_{j}\hat{p}_{j} +2\sum_{l=1}^{2} J_{jl} \hat{x}_{l} +\hat{\mathcal{N}}_{j},
\end{eqnarray}
for the mechanical dynamics.
Here
\begin{equation}
J_{jl} =\sum_{i=1}^{2}\frac{\kappa_{i} \mathrm{Im}\{G_{ij}G_{il}^{*}\} +\Delta_{i}\mathrm{Re}\{G_{ij}G_{il}^{*}\}}{(\kappa_{i}^{2}+\Delta_{i}^{2})^{2}}
\label{mecmec}
\end{equation}
is the mechanical--mechanical coupling factor and $\hat{\mathcal{N}}_{j}$ is a noise operator composed of cavity noises $\hat{X}_{i}^{\rmin}$ and $\hat{Y}_{i}^{\rmin}$ and the intrinsic mechanical noise $\hat{\xi}_{j}$.
Focusing on the interaction term reveals that the effective interaction Hamiltonian is $\sum_{j,l} \hbar J_{jl} \hat{x}_{j}\hat{x}_{l}$.
It is useful to write it in terms of phonon annihilation and creation operators $\hat{b}_{j}$ and $\hat{b}_{j}^{\dagger}$.
Then the effective Hamiltonian of the system in a frame rotating at the mechanical frequencies (or equivalently in the interaction picture with respect to $\sum_{j=1}^{2}\hbar\tilde{\Omega}_{j} \hat{b}_{j}^{\dagger}\hat{b}_{j}$) is
\begin{equation}
\hat{H}_{\mathrm{eff}} = \sum_{j,l} \hbar J_{jl} \bigg[\hat{b}_{j}^{\dagger}\hat{b}_{l}^{\dagger} e ^{ i (\tilde{\Omega}_{j}+\tilde{\Omega}_{l})t} +\hat{b}_{j}^{\dagger}\hat{b}_{l} e ^{ i (\tilde{\Omega}_{j}-\tilde{\Omega}_{l})t}\bigg] +h.c.
\label{eff}
\end{equation}
We notice that the effective Hamiltonian is composed of the following parts: (i) frequency modification of each mechanical oscillator ($\hat{b}_{j}^{\dagger}\hat{b}_{j}$), (ii) single-mode squeezing ($\hat{b}_{j}^{2} e ^{-2 i \tilde{\Omega}_{j}t} +h.c.$), (iii) phonon hopping ($\hat{b}_{1}\hat{b}_{2}^{\dagger} e ^{- i (\tilde{\Omega}_{1}-\tilde{\Omega}_{2})t} +h.c.$), and (iv) two-mode squeezing ($\hat{b}_{1}\hat{b}_{2} e ^{- i (\tilde{\Omega}_{1}+\tilde{\Omega}_{2})t} +h.c.$).
For our interest, the two-mode squeezing phenomenon is the process which can produce an entangled state of the two mechanical resonators.
However, for a CW input laser the single- and two-mode squeezing processes rapidly oscillate compared to the two remaining phenomena and have negligible effect on dynamics of the system, i.e., the rotating wave approximation allows to ignore such terms.

In order to bring the separable steady state into an entangled state one needs to excite this two-mode squeezing process.
This can be done by parametrically driving the interaction of two harmonic oscillators \cite{Mari2009,Galve2010,Farace2012}.
Generally, the mechanical--mechanical coupling in (\ref{mecmec}) can become time dependent as a result of time dependent optomechanical couplings.
This, in turn, is achieved by driving the cavity field with a pulsed or modulated input laser.
Thus, to parametrically drive the oscillators we consider a driving laser such that the intracavity field amplitude is $E_{i}(t) =E_{i}^{(0)}+E_{i}^{(1)}\cos(\omega_{\rmD}t)$, and we demand for $E_{i}^{(0)} > E_{i}^{(1)}$ to ensure that the stability conditions are not affected by the time dependent part.
From (\ref{firstm}) and the definition of $G_{ij}$ we see that for small modulation amplitudes $G_{ij}(t) \approx G_{ij}^{(0)} +G_{ij}^{(1)}\cos(\omega_{\rmD}t)$ with $G_{ij}^{(0)} > G_{ij}^{(1)}$ and conclude that
\begin{equation}
J_{jl}(t) \approx J_{jl}^{(0)} +J_{jl}^{(1)}\cos(\omega_{\rmD}t) +J_{jl}^{(2)}\cos(2\omega_{\rmD}t).
\end{equation}
Putting this in (\ref{eff}) reveals that by modulating the input lasers at $\omega_{\rmD}=\tilde{\Omega}_{1}+\tilde{\Omega}_{2}$ and $\omega_{\rmD}=(\tilde{\Omega}_{1}+\tilde{\Omega}_{2})/2$ the two-mode squeezing process will be enabled and a more efficient generation of entanglement can be expected.
Yet, since $J_{jl}^{(1)} > J_{jl}^{(2)}$ one may expect more efficient production of entanglement by modulating at sum frequency of the mechanical oscillators.

\subsection{Numerical results}
We now employ this modulated input laser in the linearized Langevin equations (\ref{secondm}) to numerically investigate the entanglement properties of the system.
Here, no more approximations are being made beyond the Lamb--Dicke approximation and linearization of the operators around their \textit{coherent} part.

\begin{figure}[tbs]
\centering
\includegraphics[width=0.9\linewidth]{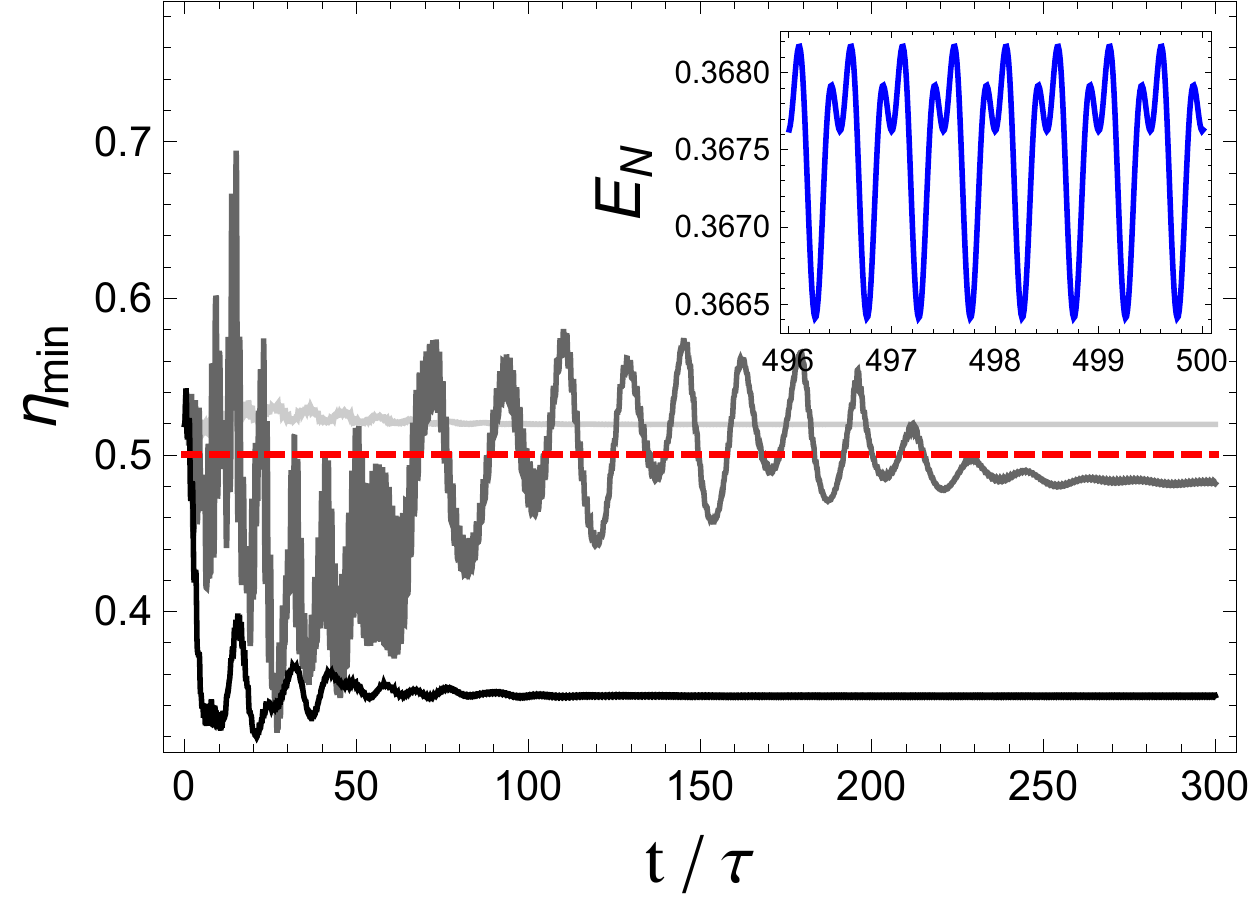}
\caption{(color online).
Evolution of minimum symplectic eigenvalue of the partially transposed mechanical--mechanical covariance matrix of two tethered microdisks optically trapped in a cavity; without modulation (light gray), modulated at average of the mechanical frequencies $\omega_{\rmD}=\frac{1}{2}(\Omega_{1}+\Omega_{2})$ (dark gray), and modulated at the sum of mechanical frequencies $\omega_{\rmD}=\Omega_{1}+\Omega_{2}$ (black).
The time is normalized to $\tau = \frac{4\pi}{\Omega_{1}+\Omega_{2}}$.
The red dashed line indicates separability criterion.
The inset shows logarithmic negativity of the mechanical entanglement at their \emph{quasi}-steady state when the cavity field is modulated at the sum frequency.}
\label{micromd}
\end{figure}

The entangling protocol we explore is the following:
The system starts with trapping the dielectric microdisks in two \emph{appropriate} antinodes---the ones which give the optimal phases---of the trapping mode which is assisted by auxiliary trapping methods for providing a precooled mechanical motion and a stable 3D optical trap.
We now turn on the CW control modes $E_{i}^{(0)}$ for further cooling down the motion of the objects close to their ground state and get close enough to the inseparability threshold (see Fig.~\ref{micross}).
Then the modulated laser beams are injected into the cavity $E_{i}^{(1)}\cos(\omega_{\rmD}t)$ to make an entangled state.
In Fig.~\ref{micromd} the results of such a procedure are summarized where the minimum symplectic eigenvalue of the bipartite full mechanical subsystem of the dielectric microdisks is plotted for two important driving frequencies: sum of the mechanical frequency ($\omega_{\rmD}=\Omega_{1}+\Omega_{2}$) and their average ($\omega_{\rmD}=\frac{\Omega_{1}+\Omega_{2}}{2}$).
The plot shows time evolutions of $\etamin$, where we have chosen the steady states resulting from CW driving amplitudes $E_{i}^{(0)}$ as the initial conditions for the evolution under modulated driving.
In the inset the logarithmic negativity of the mechanical state is shown for the case where $\omega_{\rmD}=\Omega_{1}+\Omega_{2}$ and for times where the system has well approached its asymptotic quasi-stationary regime.
One observes that $\etamin$ can be reduced below values of $\etamin=0.5$, signaling entanglement, by modulating the amplitude of the input laser.
The parameters used in the plot are the same as the parameters used in Fig.~\ref{micross} and listed in Table~\ref{parameters}.
The detunings are chosen according to the optimal steady state values $\Delta_{1} \approx \Delta_{2} \approx \Omega_{1}$, the CW intracavity amplitudes are $E_{1}^{(0)}=E_{2}^{(0)}=0.1E_{0}$, while we only modulate one of the input lasers $E_{2}^{(1)}=0.09E_{0}$.

We have here chosen the parameters such that the system is kept far from any instabilities as predicted by the Routh--Hurwitz criterion applied in section \ref{sec:steady-state}.
For this reason, the parameters that maximize the entanglement are very close to the parameters for optimal cooling and the mean phonon numbers of the microdisks show a behavior very similar to what we found for constant drive amplitudes in Fig.~\ref{micross}.

\subsection{Measuring the entanglement}
Finally, let us briefly discuss an experimental measurement method for verifying the generated entanglement.
The method we will discuss here is somewhat similar to that of Refs.~\cite{Vitali2007,Hartmann2008}.
The measurement can be done via two additional optical cavity modes.
We choose the wavelength of these cavity modes such that one of them sustains both dielectric objects at its nodes ($+$ mode) and the other mode has one object at its node while the second will lies in one of its antinodes ($-$ mode).
This will lead to a quadratic optomechanical coupling between the probe modes and the optically trapped objects.
Such a relatively weak interaction enables us to read off the state of the mechanical oscillations without making a considerable influence on the system.
In fact, its only side-effect is a modification of the stiffness of the trap which can be easily taken into account.

The equation describing the dynamics of the probe cavity modes in a frame rotating at the their resonance frequencies reads
\begin{equation}
\dot{\hat{a}}_{\pm} =-\kappa\hat{a}_{\pm} - i \sqrt{2}\tilde{\mathcal{G}}_{\pm}^{\mathrm{q}}(\mean{x}_{1}\hat{x}_{1}\pm\mean{x}_{2}\hat{x}_{2}) e ^{ i \Delta_{\pm}t} +\sqrt{2\kappa}\hat{a}_{\pm}^{\rmin},
\end{equation}
where $\tilde{\mathcal{G}}_{\pm}^{\mathrm{q}} =\sqrt{2}\mean{a}_{\pm}\mathcal{G}_{\pm}^{\mathrm{q}}$.
In the above equation we have assumed the same decay rates for both cavity modes and that both trapped objects have the same shape,
so that $\mathcal{G}_{i}^{\mathrm{q}}=\mathcal{G}_{i1}^{\mathrm{q}} \approx |\mathcal{G}_{i2}^{\mathrm{q}}|$.
By expressing the mechanical positions in terms of $\hat{x}_{j}=(\hat{b}_{j}+\hat{b}_{j}^{\dagger})/\sqrt{2}$ and further moving to the frame rotating at $\Omega = \Omega_{1} \approx \Omega_{2}$ we arrive at
\begin{eqnarray}
\dot{\hat{a}}_{\pm} &=&-\kappa\hat{a}_{\pm} - i \tilde{\mathcal{G}}_{\pm}^{\mathrm{q}}\big[(\mean{x}_{1}\hat{b}_{1}\pm\mean{x}_{2}\hat{b}_{2}) e ^{- i (\Omega-\Delta_{\pm})t} \nonumber\\
&&+(\mean{x}_{1}\hat{b}_{1}^{\dagger}\pm\mean{x}_{2}\hat{b}_{2}^{\dagger}) e ^{ i (\Omega+\Delta_{\pm})t}\big] +\sqrt{2\kappa}\hat{a}_{\pm}^{\rmin}.
\end{eqnarray}
Now we set $\Delta_{\pm}=\pm\Omega$ and drop rapidly oscillating terms to get
\begin{subequations}
\begin{eqnarray}
\dot{\hat{a}}_{+} &=&-\kappa\hat{a}_{+} - i \tilde{\mathcal{G}}_{+}^{\mathrm{q}}(\mean{x}_{1}\hat{b}_{1}+\mean{x}_{2}\hat{b}_{2}) +\sqrt{2\kappa}\hat{a}_{+}^{\rmin}, \\
\dot{\hat{a}}_{-} &=&-\kappa\hat{a}_{-} - i \tilde{\mathcal{G}}_{-}^{\mathrm{q}}(\mean{x}_{1}\hat{b}_{1}^{\dagger} -\mean{x}_{2}\hat{b}_{2}^{\dagger}) +\sqrt{2\kappa}\hat{a}_{-}^{\rmin}.
\end{eqnarray}
\end{subequations}
Typically the quadratic coupling is weak enough to have $\kappa \gg \tilde{\mathcal{G}}_{\pm}^{\mathrm{q}}$ for moderate input pumps.
Therefore, the outgoing probe cavity modes adiabatically follow the dynamics of the collective mechanical quadratures,
\begin{subequations}
\begin{eqnarray}
\hat{a}_{+}^{\mathrm{out}} &=& - i \sqrt{\frac{2}{\kappa}}\tilde{\mathcal{G}}_{+}^{\mathrm{q}}(\mean{x}_{1}\hat{b}_{1}+\mean{x}_{2}\hat{b}_{2}) +\hat{a}_{+}^{\rmin}, \\
\hat{a}_{-}^{\mathrm{out}} &=& - i \sqrt{\frac{2}{\kappa}}\tilde{\mathcal{G}}_{-}^{\mathrm{q}}(\mean{x}_{1}\hat{b}_{1}^{\dagger} -\mean{x}_{2}\hat{b}_{2}^{\dagger}) +\hat{a}_{-}^{\rmin},
\end{eqnarray}
\end{subequations}
where we have used the standard input-output relation $\hat{a}^{\mathrm{out}}=\sqrt{2\kappa}\hat{a} -\hat{a}^{\rmin}$ \cite{Gardiner2004}.
By carrying out homodyne measurements on these output modes one determines all elements of the mechanical--mechanical covariance matrix and quantifies the entanglement of the system.

%
%
\section{Levitated nanospheres}\label{sec:nanospheres}
In this section we consider the setup of a cavity with two levitated nanospheres and discuss its suitability for generating entanglement between their mechanical oscillations.
The model used in previous sections is general enough to hold for nanospheres as well.
Any trapping method such as cavity trapping \cite{Kiesel2013}, optical tweezers \cite{Millen2014}, and feedback traps \cite{Li2011} can be used in this case.
In practice, the three cavity modes---with Gaussian profile---considered in our 1D model are not enough for trapping the objects. Actually, in a 3D cavity trapping setup one needs to excite extra cavity modes with non-Gaussian profiles \cite{Yin2011} and/or external optical tweezers \cite{Millen2014} to stabilize the trap and achieve cooling in all three dimensions.

\begin{figure}[tbs]
\centering
\includegraphics[width=0.9\linewidth]{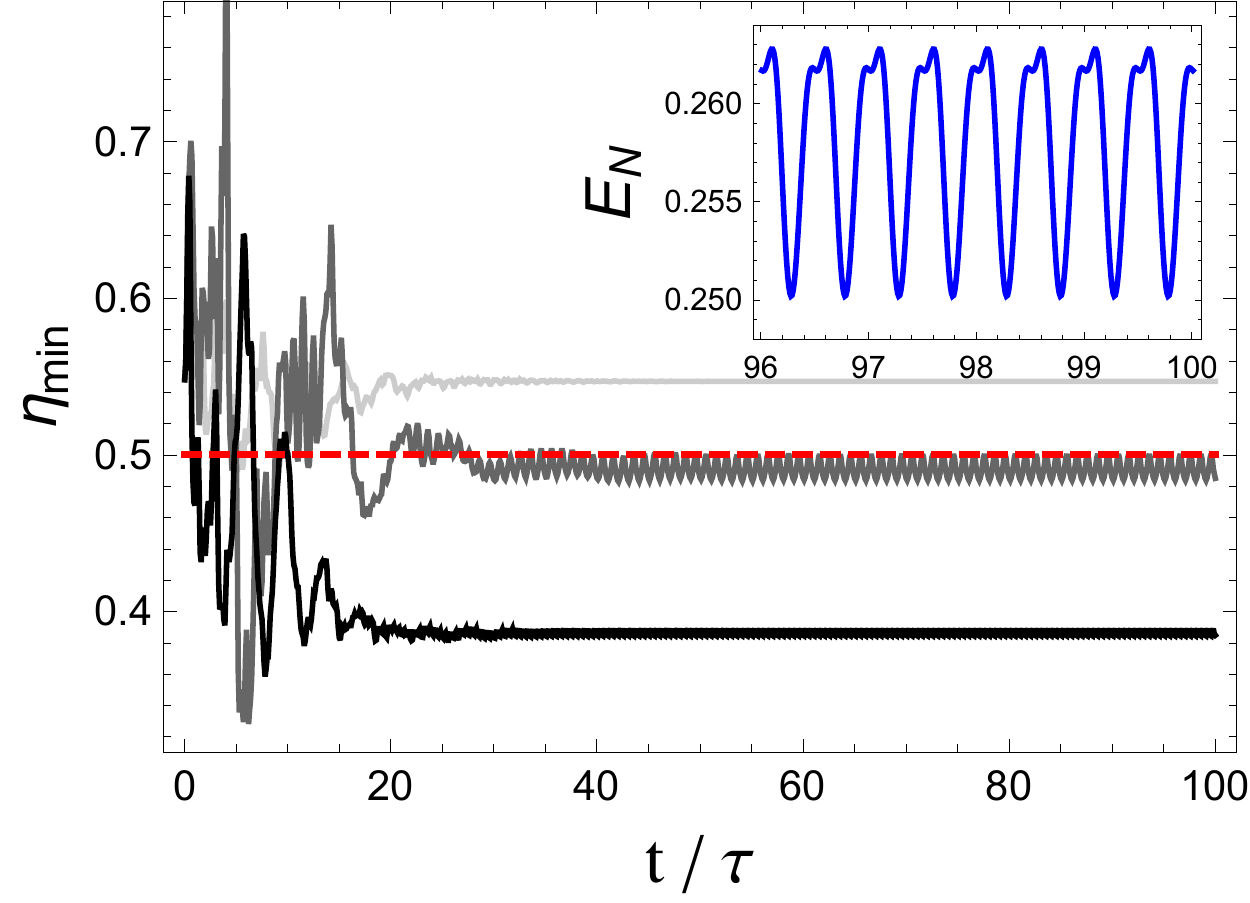}
\caption{(color online).
The same as Fig.~\ref{micromd} for two silica levitated nanospheres with equal radii of $100~\mathrm{nm}$.
The cavity detunings are $\Delta_{1} \approx \Delta_{2} \approx \Omega_{1}$, the CW and modulated intracavity amplitudes are $E_{1}^{(0)}=E_{2}^{(0)}=0.1E_{0}$ and $E_{2}^{(1)}=0.09E_{0}$, respectively.
The other parameters are listed in Table~\ref{parameters}.}
\label{nanomd}
\end{figure}

The Hamiltonian of the system is the same as (\ref{hamiltonian}) and the only required adjustment is to set coupling factors to $g_{ij}=\frac{3V_{\rmd,j}}{2V_{\rmc,i}}(\frac{\epsilon-1}{\epsilon+2})\omega_{i}$ \cite{Chang2010}.
In the case of levitated nanospheres one source of damping are collisions with residual gas molecules in the chamber, which in principle can be suppressed by lowering the chamber pressure.
However, there are some practical difficulties preventing the attainment of arbitrarily low chamber pressures \cite{Millen2014}.
Yet, recent works have shown that it is possible to get stably trapped nanospheres even at chamber pressures as low as $10^{-6}~\mathrm{mbar}$ by combining electrical and optical traps or by employing feedback mechanisms \cite{Gieseler2012}.
Despite these efforts, the coherence of the mechanical motion is here mostly affected by fluctuations in the optical trap \cite{Grotz2006} stemming from scattering of the trapping photons by the dielectric nanospheres \cite{Chang2010,Romero-Isart2011,Pflanzer2012}.
Neglecting scattering of the control mode photons, the explicit relation of $\Gamma_{j}$ in (\ref{corxi}) for nanospheres is
\begin{equation}
\Gamma_{j} = \frac{2\pi^{2}}{5}(\frac{\epsilon -1}{\epsilon +2})\frac{V_{\rmd,j}}{\lambda_{0}^{3}}\Omega_{j}.
\label{gamma}
\end{equation}
For nanospheres located inside a Fabry-P\'erot cavity the photon recoil heating is so destructive that no entanglement can  seen for the parameters of the usual experimental setups even by employing a parametric coupling.
However, the disruptive effect of the photon recoil in a nanosphere setup could be moderated by adopting a cavity with extremely close concave mirrors, since due to the almost spherical symmetry of the cavity, most of the incident photons will be prevented from scattering into free space modes, which results in a much smaller decoherence rate.
Therefore, we assume that it is possible to reduce the decoherence rate of the nanospheres to $10\%$ of its actual values, e.g. by covering at least $90\%$ of the solid angle around the trapped nanospheres with the cavity mirrors.
Fig.~\ref{nanomd} shows the possible mechanical entanglement between two $200~\mathrm{nm}$ diameter silica nanospheres, provided value of $\Gamma_{j}$ is scaled to one-tenth of that in (\ref{gamma}).
The other parameters are the same as the tethered membrane setup, as listed in Table~\ref{parameters}.

The basic problem of levitated nanospheres which hinders the creation of entangled states can be understood from the relations of $g_{ij}$ and $\Gamma_{j}$ which are both in direct proportion to the volume of the objects $V_{\rmd,j}$.
To reduce the photon recoil decoherence rate one could reduce the size of the nanosphere, which however leads to a smaller single photon optomechanical coupling.
The small single-photon coupling, however, cannot be compensated by intensely driving the cavity control modes as this will wash out the stable trapping which occurs provided the Lamb--Dicke approximation is valid.
Furthermore, even outside the Lamb--Dicke regime, for all reasonable parameter regimes, increasing the driving field intensities sufficiently to create entanglement would inevitably imply kicking the nanospheres out of the trap.

%
%
\section{Summary and conclusion}\label{sec:conclusion}
In summary, we have studied an entangling protocol for two optically trapped dielectric microdisks inside a single cavity.
We have presented a cavity trapping and controlling scheme which is also compatible with optical tweezers and feedback trapping mechanisms.
In our scheme, optical control modes provide linear optomechanical couplings which effectively lead to a mechanical--mechanical coupling.
While decoherence induced by photon recoil heating typically precludes the generation of steady state entanglement for input fields of constant intensity,
we have here shown that it is possible to push the system into an inseparable state by modulating the input lasers at proper frequencies to turn on a parametric coupling between the mechanical oscillators.
The results show a reasonable quasi-stationary mechanical--mechanical entanglement for experimentally feasible parameters.
We have also shortly discussed a possible method for measuring such a entangled state and commented on a possible setup for generating an entangled state of two levitated nanospheres.

%
%
\section*{Acknowledgment}
M.A. acknowledges support from the Alexander von Humboldt Foundation.

%
%


%
%
\bibliography{levitated}

\end{document}